\documentclass[12pt]{article}
\usepackage{epsfig,rotating,color}
\usepackage{amsmath}
\usepackage{hhline}
\usepackage{amssymb}
\usepackage{times}

\newlength{\dinwidth}
\newlength{\dinmargin}
\newcommand{\PO}{I\!\!P}

\begin{document} 
\pagestyle{empty}
\noindent
\vspace*{2cm}

\begin{center}
  \Large {\bf Experimental study of Pomeron\\}
  \vspace*{1cm} {\large A.Rostovtsev\\} 
  \vspace*{1cm} {\large \it Institute for Theoretical and Experimental
Physics\\
        Moscow, Russia } 
\end{center}
\begin{abstract}
A Pomeron phenomenon remains a mystery. A short review of the experimental
situation in diffractive physics and an account of some spectacular
manifestations of the Pomeron are given.

\noindent

\end{abstract}

\vspace*{1cm}

\newpage
\setcounter{page}{1} 
\pagestyle{plain}

\section{Introduction} \label{sect:intro}

The notion of Pomeron\footnote{Historically, it was first called
``Pomeranchuk" Regge pole.} has been introduced (Frautschi~1962) to term 
an unusual Regge pole suggested first by V.N.Gribov (1961) to describe
main properties of particle scattering at asymptotically high energies.
With advance of QCD in 70th the Pomeron phenomenology was vastly abandoned
and the emphasis has been more on perturbative QCD effects. 
During the last decade the phenomenon of the Pomeron became
again a hot subject both for theoretical and experimental scrutiny. This
increased interest is mainly generated by data from Tevatron and
HERA linking the Pomeron induced processes and the perturbative effects. 
These data and the success of QCD are important building
blocks for construction of a microscopic theory of the Pomeron. 
It is believed, that such theory should also describe the confinement
and QCD vacuum structure. This challenging task
is far from being completed now and the present theoretical understanding
the Pomeron is wired and contradictory. Below, I restrict myself to
an account of the present experimental situation and some, to my view,
spectacular experimental manifestations of the Pomeron.
It is a particular honour for me to give this talk at ITEP, in the very
building where
I.Ya~Pomeranchuk (the Russian physicist after whom the Pomeron was
termed) has being working for a long time. 

What do we (experimentalists) mean today by the ``Pomeron"? The Pomeron is 
thought to be
a colourless and flavourless extended object similar to a
hadronic resonance.  The Pomeron-hadron similarity
can be illustrated by comparison of properties of an elastic
scattering reaction, say
\begin{equation}
   p + p \rightarrow p + p\,, 
\label{eq_elastic}
\end{equation}
and a reaction with charge exchange
\begin{equation}
   \pi^- + p \rightarrow \pi^0 + n\,. 
\label{eq_ce}
\end{equation}
In both reactions the colliding particles exchange something 
which carries a 4-momentum and specific quantum numbers.
Regge theory says, that a cross section of these reactions is a simple
function  
\begin{eqnarray*}
   \frac{d\sigma}{dt} \sim s^{2\alpha(t)-2}\,, 
\end{eqnarray*}
where $s$ and $t$ are total energy and momentum transfer squared.
 Figure~1 shows the measurements (Bolotov~1974,
Akerlof~1976) of the
$\alpha$ as function of $t$ for reactions~(\ref{eq_elastic}) 
and~(\ref{eq_ce}). These measurements
can be parameterized by a linear form, called {\it trajectory}
\begin{eqnarray*}
   \alpha(t) = \alpha(0) + \alpha' t\,. 
\end{eqnarray*}
In case of the reaction~(\ref{eq_ce}) the $\rho$ is a resonance with
correct quantum numbers exchanged in the reaction. If the spin of
$\rho$ is plotted against the square of its mass on 
Fig.~1a one finds
that extrapolation of the trajectory to positive values $t\equiv{M^2}$
goes through the $\rho$ point. Moreover, there is a rich family of
resonances, which belong to this trajectory. This is a beautiful
finding related to the analyticity of scattering amplitudes.

The trajectory for elastic reaction has parameters which differ from
those of $\rho-$trajectory. Namely, an intercept 
$\alpha_{\PO}(0)\approx~1.1$ and a slope $\alpha'_{\PO}\approx
0.25~GeV^{-2}$.
This trajectory is called Pomeron trajectory and the object exchanged
in the elastic reaction~(\ref{eq_elastic}) is called Pomeron. It carries
vacuum quantum
numbers (C=P=+1), and is thought to be like a gluonic bubble. Though, 
the Pomeron
trajectory has no undebatable particles on it, it is widely agreed that,
if they will be found, they are glueballs.
The Pomeron trajectory plays an important role at high energies describing
an universal rise of the total $pp, \pi{p}, Kp, \gamma{p}$ cross
sections~(Donnachie 1992) 
and a shrinkage of the forward peak in diffractive 
processes.

Since the Pomeron is
a colourless object it's signature in a high energy event is a void of
particles spanning
a large rapidity interval, so-called {\it rapidity gap}. 
Unlike fluctuations
in gluon fragmentation the Pomeron induced rapidity gaps are not
exponentially suppressed. A class of reactions with non-exponentially
suppressed large rapidity gaps is operationally termed diffractive
reactions~(Bjorken~1994). 
It is convenient to classify the diffractive reactions by a position
of large rapidity gap in the event.
In diffractive reactions with a beam hadron staying intact
the rapidity gap is adjacent to the direction of this beam particle.
The reaction with rapidity gap adjacent to only one of the beams
is termed {\it single diffractive dissociation~(SDD)}. The inelastic 
reaction with two rapidity gaps adjacent to the both beams
is called {\it Double Pomeron Exchange~(DPE)}.
The reaction with both
beam particles dissociating in to hadronic systems
and leaving a rapidity gap inbetween is called
{\it double diffractive dissociation~(DDD)}. 

An example of a measured by H1~(1995) rapidity gap width distribution is
shown in Figure~2 in terms of $\eta_{max}$ variable. With decreasing
of the $\eta_{max}$ the rapidity gap grows.
One observes the exponentially falling spectrum flattens out for
$\eta_{max}<2$. In this region of large rapidity gaps the
diffractive process dominates and can be experimentally separated from
non-diffractive events.
Why the rapidity gap spectrum for diffractive reaction is
flat? The answer to this question points to a radiative type of the
Pomeron emission by beam particles similar to radiation of photon.
For experimentalists the picture suggested by Ingelman and
Schlein~(1985), is very illustrative. First, a
Pomeron is emitted from one of beam particles with momentum transfer~$t$
and probability
$P \sim 1/x_{\PO}$, where $x_{\PO}$ is a fraction of the beam energy taken
by the Pomeron. Then, this Pomeron interacts
inelastically with the other incoming beam particle. Assume, the
Pomeron has typical hadronic properties, so that the system $X$, produced
in a Pomeron-particle collision, would resemble normal minimum bias events
in hadron-hadron inelastic scattering at corresponding invariant mass
$M_X^2=x_{\PO}(2E_{beam})^2$. This system decays in to final state hadrons
with momentum distribution given by a longitudinal phase space. 
A rapidity interval filled by the final
state hadrons is $\Delta\eta \sim log(M_X^2)$, then  
\begin{eqnarray*}
  \frac{d\sigma_{diff}}{d\Delta\eta} \sim 
   M_X^2\frac{d\sigma_{diff}}{dM_X^2} \sim x_{\PO}\frac{1}{x_{\PO}}\sim
const \,.
\end{eqnarray*}
 In a fraction of events resulting from the Pomeron-particle interaction
high $P_t$ hadronic jets are produced. In the Ingelman-Schlein model
the produced jets carry a direct information on the partonic content
of the Pomeron. Following this approach, the DPE reaction is seen like
a collision of two Pomerons. Interesting to note, though the
resulting hadronic
system is produced fully isolated in a rapidity space, and nevertheless, 
it is aligned with the initial beam direction.  

The above described picture of diffraction assumes a factorization of
Pomeron flux and the Pomeron-particle cross section. In order to make
this picture quantitative it is practical to use the Pomeron flux predicted
by the Regge theory.  
In a framework of Regge theory the cross section for a single
diffractive dissociation resulting from one Pomeron exchange
can be written (see e.g. Goulianos~1983) as
\begin{equation}
   \frac{d^2\sigma_{SDD}}{dx_{\PO}dt} =
[\frac{\beta^2(t)}{16\pi x_{\PO}^{2\alpha_{\PO}(t)-1}}]\times
\{g\beta(0)(\frac{x_{\PO}s}{s_0})^{\alpha_{\PO}(0)-1}\}\,,
\label{Regge}
\end{equation}
where $\beta(t)$ is the coupling of the Pomeron to a beam hadron,
$g$ is the triple-Pomeron coupling, $s_0$ is an arbitrary scale parameter.
The function $\beta(t)$ is obtained from the elastic scattering
experiment. The term in the square brackets is interpreted as a Pomeron
flux, while the remaining factor is identified as a Pomeron-hadron total
cross section. Further on the Pomeron flux defined by~(\ref{Regge}) is 
termed ``standard Pomeron flux".
The factorization formula~(\ref{Regge}) has correct $M_X^2$ behaviour
\begin{equation}
   \frac{d^2\sigma_{SDD}}{dM_X^2dt}|_{t=0} \sim 
(\frac{1}{M_X^2})^{1.1}\,,
\label{sd}
\end{equation}
measured in the experiments~(Goulianos 1999).
At the same time, the formula~(\ref{Regge}) implies fast rise of
the diffractive cross section with energy
\begin{eqnarray*}
 \frac{d^2\sigma_{SDD}}{dM_X^2dt}|_{t=0} \sim s^{2\alpha_{\PO}(0)-2}\,,
\end{eqnarray*}
which is not supported by the experimental data discussed in the
following section.

\section{Pomeron at HERA and Tevatron}

The measurements at HERA of deep-inelastic electron-proton
scattering~(DIS) have demonstrated that about $10\%$ of events have 
large rapidity gap adjacent to the proton beam direction. 
A natural interpretation of these rapidity gap events
is based on the hypothesis that highly virtual photon probes a Pomeron
emitted from the proton. Therefore,
at HERA one can measure a partonic structure of the Pomeron the same
way as the proton structure function $F_2^p$ is measured. By analogy
with the $F_2^p$ one defines the diffractive structure function 
$F_2^{D(3)}$
\begin{eqnarray*}
\frac{d\sigma_{ep\rightarrow epX}}{dxdQ^2dx_{\PO}} =
\frac{4\pi \alpha^2}{xQ^4}(1-y+\frac{y^2}{2})F_2^{D(3)}, 
\end{eqnarray*}
where $x, Q^2$ are usual variables used to describe DIS.
The variable $x_{\PO}$ is calculated in the experiment from the measured
mass of hadronic system $M_X$ resulting from photon-Pomeron interaction
\begin{eqnarray*}
  x_{\PO} = \frac{x}{\beta}\,\,\,,\,\,\,
\beta \approx \frac{Q^2}{Q^2+M_X^2}\,,
\end{eqnarray*}
where $\beta$ 
is a ratio of the momentum carried by the quark coupling to a virtual
photon to that of the Pomeron. In another words, $\beta$ is an analog of
$x$-Bjorken for the Pomeron.

Figure~3 shows the measured by the H1 Collaboration
(1997) $F_2^{D(3)}$ as function of
$Q^2$ for different values of $\beta$ at fixed $x_{\PO}$. 
The data show a relatively flat $\beta$ dependence
and a rising dependence on $Q^2$, except at the highest values of
$\beta$. This structure is well described by a fit based on
DGLAP evolution of the $\beta$ and $Q^2$ dependence and a Regge 
motivated $x_{\PO}$ dependence. In this fit, the diffractive
parton distributions of the proton are heavily dominated by a large
gluon density. The scaling violations of $F_2^{D(3)}$ are similar to
those of $F_2$ when compared at the same $x$ values, except at large
$\beta$ where vector meson and other higher twist contributions
are expected to play a significant role in the diffractive data. 

To extract the Pomeron parton densities we have to assume a factorization
of $F_2^{D(3)}$ into a Pomeron flux and a Pomeron structure function
\begin{eqnarray*}
  F_2^{D(3)}(x_{\PO},\beta,Q^2) = f_{\PO/p}(x_{\PO})
F_2^{\PO}(\beta,Q^2)\,.
\end{eqnarray*}
The Pomeron flux is taken from formula~(\ref{Regge}) 
and integrated over $t$. Under the hypothesis of diffractive factorization,
the parton distributions for the Pomeron extracted from $F_2^{D(3)}$ are
expected to describe diffractive interactions wherever perturbation theory
may be applied. The Tevatron data allow such cross check.

At Tevatron the CDF~(2001) has measured
the diffractive structure function of the antiproton using
a method employing two samples of dijet events produced in $p\bar{p}$
collisions at $\sqrt{S}=1800~GeV$: a single proton
diffractive dissociation sample, and an inclusive sample, collected with a
minimum bias trigger. In leading order QCD, the ratio of the diffractive
to inclusive cross sections as function of Bjorken $x$ of the struck
parton in the antiproton, obtained from the jet kinematics, is equal to
the ratio of the corresponding structure functions. This method gives a
model independent colour-weighted structure function
\begin{eqnarray*}
F_{jj}^D(x)=\{g^D(x)+\frac{4}{9}\sum_i[q_i^D(x)+
\bar{q}_i^D(x)]\}\,,
\end{eqnarray*}
where $g^D(x)$ and $q_i^D(x)$ are the antiproton gluon and quark
diffractive parton densities. By changing $x$ to $\beta=x/x_{\PO}$ the
obtained parton distribution can be compared with that calculated 
using parton densities in Pomeron measured at HERA and the standard
Pomeron flux. This comparison is shown in Figure~4.
The Tevatron and HERA parton distributions disagree 
both in shape and normalization by
an order of magnitude. Similar sizable disagreement with predictions based
on the Pomeron structure measured at HERA and the standard Pomeron flux
are reported by CDF in reactions of $W$, $b$-quark and $J/\Psi$ 
diffractive production (Goulianos~2001). Thus, under the factorization
hypothesis a Pomeron at the Tevatron
has 10 times lower parton density than that at HERA! 
In addition, a steeply rising SDD cross section given
by~(\ref{Regge}) overshoots the Tevatron measurement of $\sigma_{SDD}$ by
a similar factor~(Goulianos 1995).

A way out of this contradiction is to revise 
calculations of the Pomeron flux. It was suggested (Kaidalov 2001)
that at Tevatron energies additional parton rescatterings kill rapidity
gaps and therefore, effectively suppress the standard Pomeron flux. 
In this approach a suppression factor
is different for diffractive jet production and for inclusive diffraction. 
Another approach is to renormalize a rapidity gap probability
(Goulianos 2000) in order to describe experimentally observed
scaling law for diffractive dissociation. 
A spectacular graphical representation of this scaling law is shown in
Figure~5, where the measured spectra
$d\sigma^2/dtdM_X^2$ at fixed $t$ are plotted
as function of $M_X^2$ for $pp$ collision energies from $\sqrt{s}=14$ to
$\sqrt{s}=1800~GeV$. This data show an independent of $s$
universal behaviour of the differential diffraction cross section.
This remarkable universality is highly non-trivial and disagree with
$s$ dependence expected from the Regge formula~(\ref{Regge}). 
Whatever theoretical approach is used to 
make corrections to the Regge flux it looks like a
construction of a Ptolemaic system. 
The factorization breaks down such a way that gives rise to a 
universal scaling law!

\section{Exclusive diffractive meson production}

The HERA collider provides a unique opportunity to observe a rich
family of exclusively produced vector mesons in a diffractive reaction. 
\begin{equation}
e + p \rightarrow e' + p' + V
\label{vm}
\end{equation}
In $ep$ scattering the photon irradiated by the electron interacts with 
the Pomeron emitted from proton and may materialize into a meson with
quantum numbers of the photon $(J^{CP}=1^{--})$.

In Figure~6 the cross section for elastic
electroproduction of vector
mesons is presented as function of the variable $(Q^2+M_V^2)$. The data
in Figure~6 compile the HERA measurements~(H1 2000)
at $\sqrt{s_{\gamma{p}}}=75~GeV$ scaled by SU(5) factors,
according to the quark charge content of the vector mesons, which amount
to 1 for the $\rho$, 9 for the $\omega$, 9/2 for the $\phi$ and $\Upsilon$
and 9/8 for the $J/\Psi$ meson. Within the experimental errors, the
exclusive vector meson production cross sections, including the SU(5)
normalization factors, turned out to lie on a universal curve which
is close to a simple $\sim 1/(Q^2+M_V^2)^{2.2}$ behaviour. 
Note, that that the ratio of SU(5) factors is in fact can be approximated
by ratios of the electronic width $\Gamma_{ee}$ of the vector mesons.
Thus, for the photoproduction ($Q^2\approx 0$) the dependence of the
exclusive vector meson cross section is
\begin{equation}
\sigma_{\gamma{p}\rightarrow Vp} 
\sim \Gamma_{ee}(\frac{1}{M_V^4})^{1.1}\,.
\label{vm1}
\end{equation}
The extra power term in (\ref{vm1}) and (\ref{sd}) is taken to be
$\alpha_{\PO}=1.1$, which agrees with the data within the experimental
errors.
Interesting to observe that this universality holds true for light
$\rho^0$, where perturbative calculations are not directly applicable and
for heavy quarkonia states, where it is believed that perturbative
predictions are straight forward. In addition, at $Q^2=0$ the photon
interacting with the target proton has a hadronic structure. As found
by Block~(2001), the rescattering effects are expected to suppress the
diffractive $\gamma{p}$ cross section in a same way as in hadron-hadron
collisions.
In addition these corrections work differently for low mass vector mesons
and for heavy $J/\Psi$ and $\Upsilon$.
Remarkable, that after cooking these corrections in a soup from
particles of different masses spiced with various flavours of constituent
quarks the beautiful and simple scaling law~(\ref{vm1}) emerges.

It is interesting to point out a close similarity of exclusive
diffractive vector
meson production in $\gamma{p}$ to inclusive meson production in
hadronic collisions. It was found (Gaisser~1978) that the inclusive cross 
section for hadroproduction of narrow vector mesons $(\phi, J/\Psi,
\Upsilon)$ follow the scaling law
\begin{equation}
 \sigma_{pp\rightarrow VX} \sim \frac{\Gamma_{ggg}}{M_V^4}\,,
\label{vm2}
\end{equation}
where $\Gamma_{ggg}$ is a hadronic width for the Okubo-Zweig-Iizuka rule 
violating decays of
$J/\Psi$ and $\Upsilon$, and calculated three gluon width for the $\phi$.
This scaling law spans over magnitudes of
difference in mass, width and production cross section of the mesons.
The scaling law~(\ref{vm2}) was found at that time in low energy
$pp$ collisions. I am not aware of any check that this simple relation 
works at Tevatron energy. 

As an exercise, we divide the inclusive production cross
sections by the full width of flavourless particles produced at
Tevatron energy $\sqrt{s}=1800~GeV$ and plot this ratio as function of
particles mass.
Figure~7 shows the compilation of the data, where
in order to
simplify a direct comparison, the differential production cross section
$d\sigma/dy(y=0)$ is used.
The data for $\Upsilon, W$ and $Z$ come from
CDF~(1995), D0~(1999), and CDF~(2000a) correspondingly.
The data for $\rho, \phi$ and $f_2$ are not available directly at Tevatron
and were extrapolated from the ISR measurements (Albrow~1979, Drijard~1981).
The $J/\Psi$ production cross section was 
estimated by extrapolation of measured $P_t$ spectrum (Sansoni~1996).
Wherever the extrapolation was applied, additional $80\%$ systematic
uncertainty is attributed to the data points on the plot in
Figure~7. 
Remarkably, the mesons roughly follow a scaling law similar to
(\ref{vm}) and~(\ref{vm1})
\begin{equation}
 \frac{d\sigma_{pp\rightarrow MX}}{dy}(y=0) \sim \frac{\Gamma}{M^4}\,.
\label{vm3}
\end{equation}
The narrow $J/\Psi$ and $\Upsilon$ resonances lie above the rest of the
particles, though, they follow the universal dependence~(\ref{vm3}).
Could it be pure accidental, that
the light hadrons $\rho, \phi$ and $f_2$ lie on the same line as
heavy and pointlike gauge bosons $W$ and $Z^0$?
To extend the exercise
we add to the plot in Figure~7 an area of the
expected values $\sigma/\Gamma$ for low mass Higgs boson production at the
Tevatron (Carena~2000). It turned out, that expected Higgs production
cross section is close to a naive estimate based on the scaling
law~(\ref{vm3}).

Like the mesons produced in diffractive reaction~(\ref{vm}) carry quantum
numbers of photon, the exclusively produced in DPE mesons have quantum
numbers of the Pomeron ($I^GJ^{CP}=0^+J^{++}$). Generally, the DPE
is one of the most spectacular phenomenon and might
be seen as a natural laboratory to study Pomeron-Pomeron interactions. 
For inclusive DPE reaction it was found (Breakstone~1989) 
that resulting hadronic system resembles that produced in a single
diffractive dissociation. It shows no
anomaly in relative yield of pions kaons and protons with respect to that
in non-diffractive inclusive hadroproduction, and the differential cross
section of DPE process observes $1/M_X^2$ behaviour typical for other
classes of inelastic diffractive reactions. 
Further on we will consider the exclusive DPE reactions only.

Experimentally the DPE exclusive state production has been extensively
studied in $pp$ collisions at energies up to ISR energies. 
The available data are limited to light $f_0, f_1$ and $f_2$ mesons.
Those mesons 
are produced at low $|t|$ and its production cross section rises slowly
with the collision energy~(Kirk 2000). 
A prominent example of DPE meson production is
\begin{eqnarray*}
 p + p \rightarrow p + f_2(1270) + p\,,
\end{eqnarray*}
with measured cross section at $\sqrt{s}=62~GeV$ 
to be $14\pm5~\mu{b}$ (Breakstone~1986).
The DPE cross section rises slowly with $s$ and
one could expect a twice larger cross section for exclusive $f_2(1270)$ 
production at Tevatron energy.  
The Tevatron experiments have large phase space to produce
heavy particles in the DPE reaction.
Taking a hypothesis that the exclusive DPE meson production observes
the scaling law found for inclusive hadroproduction~(\ref{vm3})
one would expect
for $\chi_c$ DPE production about $3~nb$. This cross section is large
enough to be measured at the Tevatron
provided a trigger is efficient
for low $P_t$ muons and electrons from a cascade decay.

The Higgs boson has correct quantum numbers and can be produced
exclusively in the DPE reaction
\begin{equation}
 p + p \rightarrow p + H + p\,.
\label{pHp}
\end{equation}
Presently, the theoretical predictions for the 
$pp\rightarrow{pHp}$ cross section at the Tevatron vary form about $0.1$
to $100~fb$ for the low mass Higgs.
 Such big uncertainty in the calculations reflects 
the difficulty to describe non-perturbative effects associated with
the formation of rapidity gaps. We boldly go ahead and apply the scaling
law for the exclusive Higgs production at Tevatron
\begin{eqnarray*}
     \sigma(pp\rightarrow{pHp})=
\sigma(pp\rightarrow{pf_2p})\frac{\Gamma_H M_f^4}{\Gamma_f M_H^4}
\approx 10~fb\,.
\end{eqnarray*}
If this estimate will come to be correct, the $pp\rightarrow{pHp}$
reaction give a unique chance to discover low mass Higgs at Tevatron.
Recently, it has been shown (Albrow~2000) that a measurement of
outgoing $p$ and $\bar{p}$ momenta in the reaction~(\ref{pHp})
will allow to reach unprecedented
Higgs mass resolution of $250~MeV$ by calculating a missing mass.
This challenging measurement will require a precise monitoring of the
beam parameters, probably, using high statistics elastic $p\bar{p}$
scattering. 
Given expected for Tevatron Run~II luminosity of $15~fb^{-1}$ and average
acceptance of $50\%$ the total number of reconstructed $pHp$ events
is $75$ at a negligible background. A relatively low cost of detector
modification required for this measurement makes this measurement very
attractive in the future.
The Pomeron physics was before and is now a rich source of information about
QCD vacuum and non-perturbative phenomena. I believe, it will remain an
experimental field of spectacular observations.

\vspace*{1.0cm}
{\Large\bf Acknowledgements}\\

The work was partially supported by 
Russian Foundation for Basic Research, grant \\
RFBR-01-02-16431.

\newpage

{\Large\bf References}

Akerlof, C.W., {\it et al}, ``Hadron-proton elastic scattering at 
$50, 100$ and $200~GeV/c$ momentum", Phys.Rev. {\bf D14}, 2864 (1976).

Albrow, M.G. and Rostovtsev, A., ``Searching for the Higgs boson at hadron
colliders using the missing mass method", FERMILAB-PUB-00-173,
e-Print Archive: hep-ph/0009336, (2000).

Albrow, M.G., {\it et al}, ``Inclusive $\rho^0$ production in $pp$
collisions at the CERN ISR", Nucl.Phys., {\bf B155}, 39 (1979).

Bjorken, James D., ``Hard diffraction and deep inelastic scattering",
Talk given at International Workshop on Deep
Inelastic Scattering and Related Subjects,
Eilat, Israel, 6-11 Feb 1994. 
Published in Eilat DIS Workshop, 151
(1994).

Block, M.M. and Halzen, F., ``Survival Probability of Large Rapidity Gaps
in $\bar{p}p, pp, \gamma{p}$ and $\gamma\gamma$ Collisions",
Phys.Rev. {\bf D63}, 114004 (2001).

Bolotov, V.N., {\it et al}, ``Negative pion charge exchange scattering on
protons in the momentum range $20-50~GeV/c$", Nucl.Phys. {\bf B73},
365 (1974). 

Breakstone, A., {\it et al}, ``Inclusive Pomeron-Pomeron interactions at
the CERN ISR", Z.Phys., {\bf C42}, 387 (1989).

Breakstone, A., {\it et al}, ``Production of the $f^0$ meson in the
double Pomeron exchange reaction $pp\rightarrow pp\pi^+\pi^-$ at
$\sqrt{s}=62~GeV$", Z.Phys., {\bf C31}, 185 (1986).

Carena, M., {\it et al}, ``Report of the Tevatron Higgs working group",
FERMILAB-CONF-00-279-T, e-Print Archive: hep-ph/0010338, (2000).

%
CDF  Collab., ``Measurement of $d\sigma/dy$ for high mass Drell-Yan
$e^+e^-$ pairs from $p\bar{p}$ collisions at $\sqrt{s}=1800~GeV$",
Phys.Rev., {\bf D63}, 011101 (2001).

CDF Collab., ``Diffractive dijets with a leading $\bar{p}$ in $\bar{p}p$
collisions at $\sqrt{s}=1800~GeV$", Phys.Rev.Lett. {\bf 84}, 5043,(2000).

CDF Collab., ``$\Upsilon$ production in $p\bar{p}$ collisions at
$\sqrt{s}=1800~GeV$", Phys.Rev.Lett., {\bf 75}, 4358 (1995).

D0 Collab., ``Measurement of $W$ and $Z$ boson production cross-sections",
Phys.Rev., {\bf D60}, 052003 (1999).

Drijard, D., {\it et al}, ``Production of vector and tensor mesons in
proton-proton collisions at $\sqrt{s}=52.5~GeV$", 
Z.Phys., {\bf C9}, 293 (1981).

Donnachie, A. and Landshoff, P.V., ``Total cross sections", 
Phys.Lett. {\bf B296}, 227 (1992).

Frautschi, S.C., Gell-Mann, M. and Zachariasen, F, ``Experimental
consequences of the hypothesis of Regge poles", Phys.Rev. {\bf 126},
2204 (1962).

Gaisser, T.K., Halzen, F., and Paschos and E.A., ``Hadronic production of
narrow vector mesons", Phys.Rev., {\bf D15}, 2572 (1977).

Goulianos, K., ``The Diffractive structure function at the Tevatron: CDF
results",\\
Nucl.Phys.Proc.Suppl. {\bf 99A}, 37 (2001).

Goulianos, K., ``Diffraction: a new approach", J.Phys., {\bf G26}, 716
(2000).

Goulianos, K., and Montanha, J., ``Factorization and scaling in hadronic
diffraction",\\
 Phys.Rev., {\bf D59}, 114017 (1999).

Goulianos, K., ``The structure of the pomeron", 
e-Print Archive: hep-ph/9505310,  (1995).

Goulianos, K., ``Diffractive interactions of hadrons at high energies",
Phys.Rep. {\bf 101}, 169 (1983).

Gribov, V.N., ``Possible asymptotic behavior of elastic scattering",
Soviet Phys.-JETP {\bf 14}, 478 (1961).

H1 Collab., ``Measurement of elastic electroproduction of $\phi$
mesons at HERA", Phys.Lett., {\bf B483}, 360 (2000).

H1 Collab., ``Inclusive measurement of diffractive deep-inelastic
ep scattering", Z.Phys. {\bf C76}, 613 (1997).

H1 Collab., ``Observation of hard processes in rapidity gap events in
gamma-p interactions at HERA", Nucl.Phys., {\bf B435}, 3 (1995).

Ingelman, G. and Schlein, P., ``Jet structure in high mass diffractive
scattering", Phys.Lett. {\bf 152B}, 256 (1985).

Kaidalov, A.B., Khoze, V.A. Martin, A.D. and Ryskin, M.G.,
``Probabilities of rapidity gaps in high energy interactions",
e-Print Archive: hep-ph/0105145, (2001).

Kirk, A., ``Resonance production in central $pp$ collisions
at the CERN Omega Spectrometer", Phys.Lett., {\bf B489}, 29 (2000).

Newman, P. ``Measurement of the diffractive structure function at HERA",
Talk given at the International Europhysics Conference on High Energy
Physics, Budapest, July (2001).

Sansoni, A., ``Quarkonia production at CDF", Talk given at
12th International Conference
on Ultra-Relativistic Nucleus-Nucleus
Collisions (Quark Matter 96), Heidelberg,
Germany, 20-24 May 1996.,
Published in Nucl.Phys., {\bf A610} 373, (1996).

\newpage

{\Large\bf Figures}

{\bf Figure 1:} Regge trajectories for the charge exchange~(a) and
elastic scattering~(b) reactions.

{\bf Figure 2:} A spectrum of the rapidity gap width distribution measured
at H1~(1995).

{\bf Figure 3:} Dependence of $F_2^{D(3)}$ on
$Q^2$ for different values of $\beta$ at fixed $x_{\PO}$. The data are
taken from~(Newman~2001).

{\bf Figure 4:} A comparison of the colour-weighted Pomeron parton
densities as measured at HERA~(upper curves) and at Tevatron~(the data
points below).

{\bf Figure 5:} Double differential SDD cross section for 
$p(\bar{p})+p\rightarrow p(\bar{p})+X$ at $t=-0.05~GeV^2$ and
$\sqrt{s}=14, 20, 546$ and $1800~GeV$.

{\bf Figure 6:} A Compilation of HERA measurements of the total cross
sections $\sigma(\gamma^*p\rightarrow Vp)$ as function of $(Q^2+M_V^2)$
for elastic $\rho, \omega, \phi, J/\Psi$ and $\Upsilon$ meson production
at the fixed $\sqrt{s_{\gamma^*p}}\equiv W=75~GeV$. The measured cross
sections were scaled by SU(5) factors according to the quark charge
content of the vector meson. The insert presents a ratio of the scaled
$\omega, \phi$ and $J/\Psi$ cross sections to the parameterization curve.
 
{\bf Figure 7:} The $d\sigma/dy(y=0)$ cross section scaled by the values
of the full width as function of the particle mass
for $\rho, \phi, f_2(1270), J/\Psi, \Upsilon, W^+$ and
$Z$ particles produced in $pp$ collisions at $\sqrt{s}=1800~GeV$.
The curves illustrate $1/M^4$ dependence. The area of
expected values $\sigma/\Gamma$ for low mass Higgs boson production at
Tevatron is shown with a large filled circle.

\newpage
\begin{figure}[ht]
\vspace*{-0.5cm}  
\begin{center}
\epsfig{
file=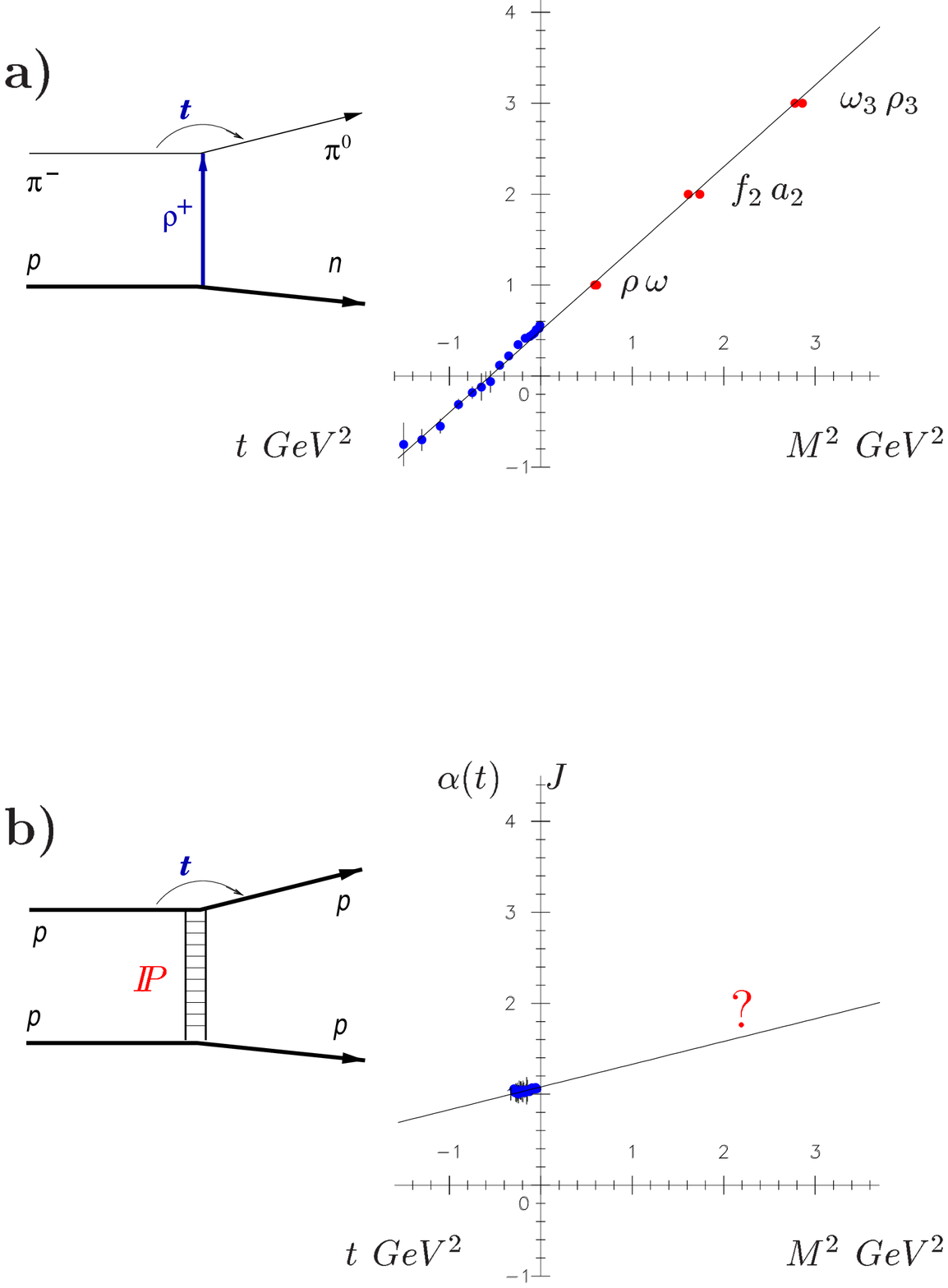,
        height=25.0cm,angle=0}
\put(-110,20){\bf Figure~1}
\end{center}
\label{Figure-Regge}
\end{figure}
\newpage
\begin{figure}[ht]
\begin{center}
\hspace*{-2.0cm}
\epsfig{
file=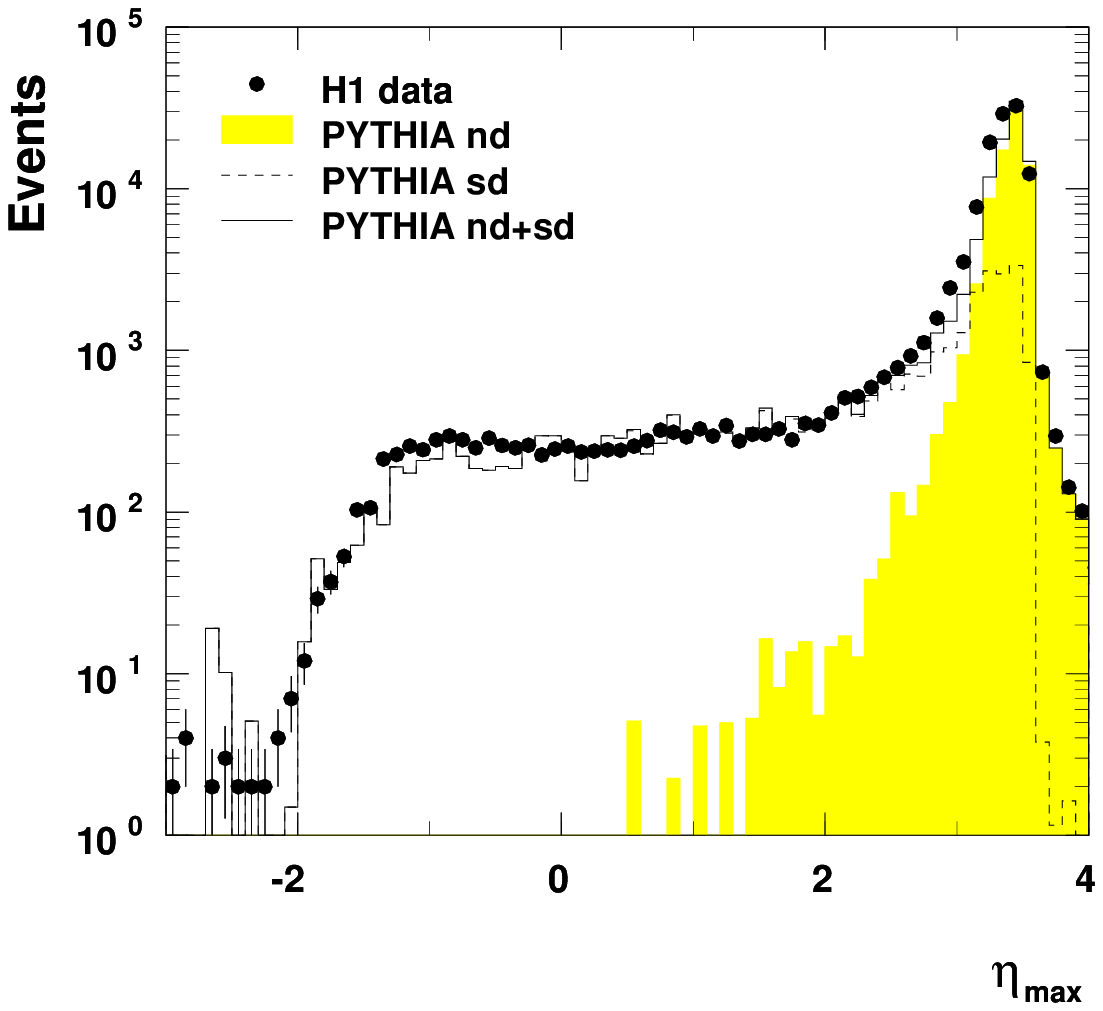,
        bbllx=100,bblly=480,bburx=480,bbury=800,clip=,
        height=15.0cm,angle=0}
\put(-95,-35){\bf Figure~2}
\end{center}
\label{Figure-gap}
\end{figure}
\newpage
\begin{figure}[ht]
\begin{center}
\hspace*{-2.0cm}
\epsfig{
file=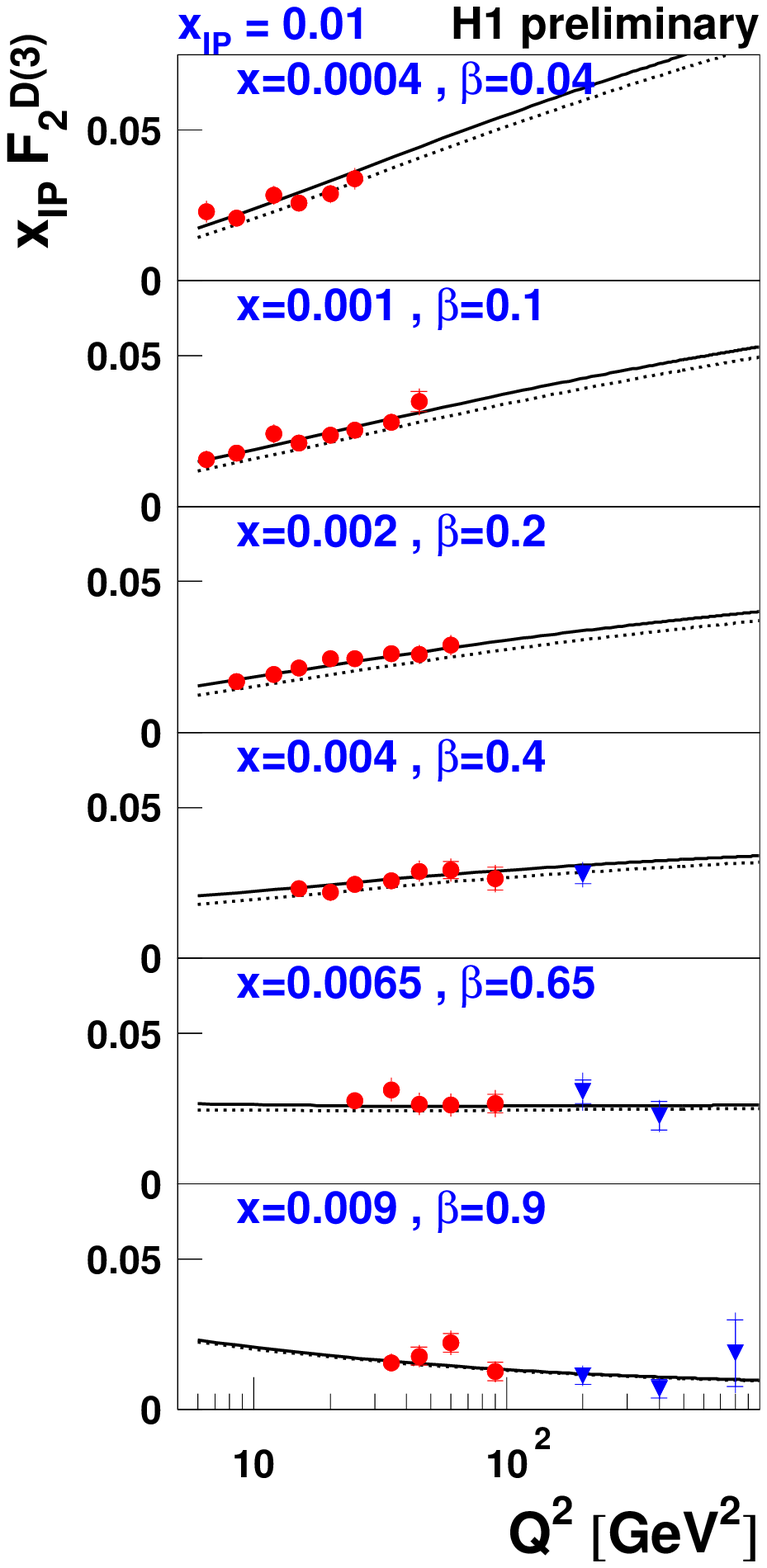,
        height=20.0cm,angle=0}
\put(-55,0){\bf Figure~3}
\end{center}
\label{Figure-pomstf-q2}
\end{figure}
\newpage
\begin{figure}[ht]
\begin{center}
\hspace*{-2.0cm}
\epsfig{
file=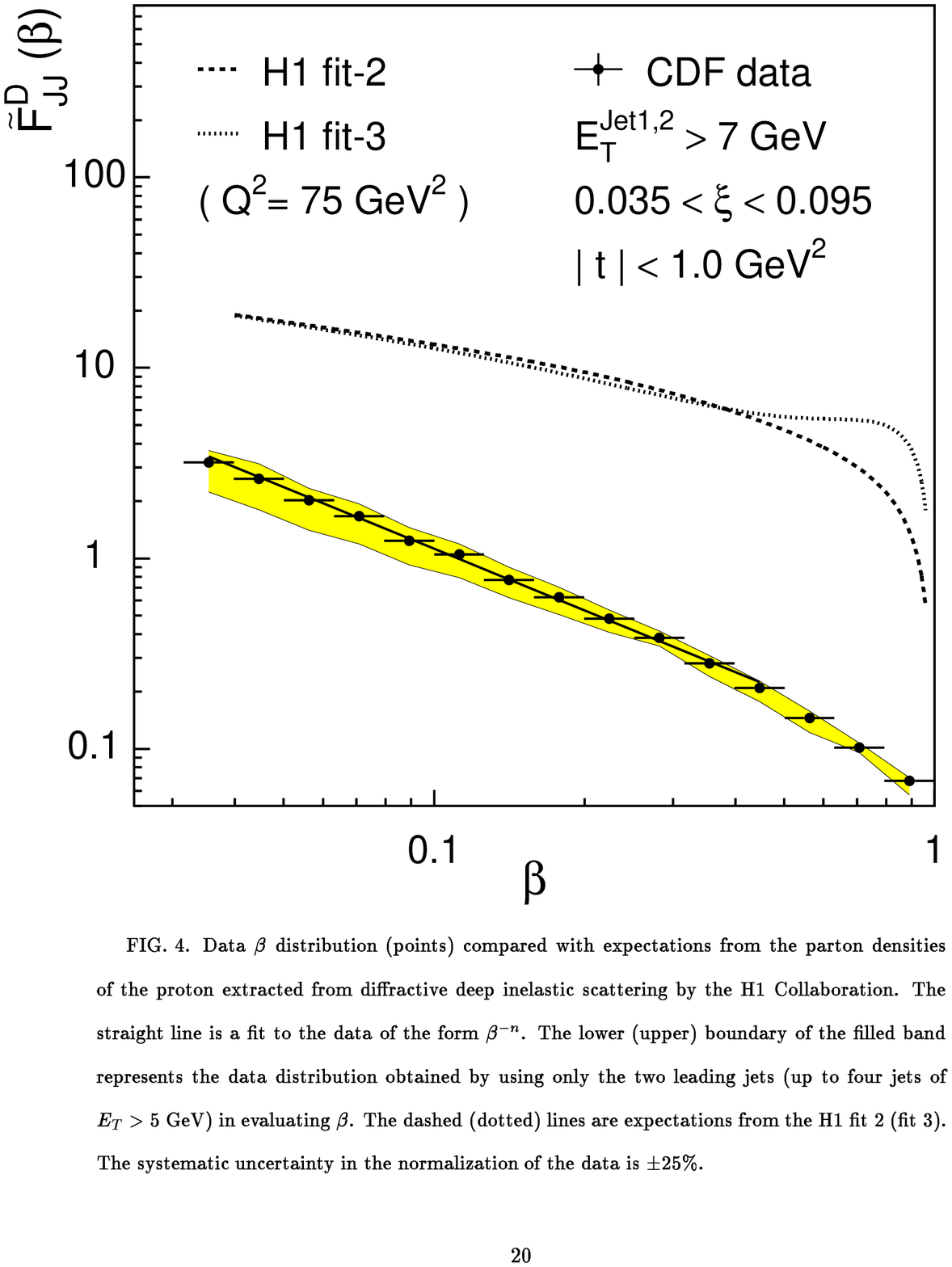,
        bbllx=0,bblly=220,bburx=580,bbury=790,clip=,
        height=15.0cm,angle=0}
\put(-75,-35){\bf Figure~4}
\end{center}
\label{Figure-pomstf-cdf}
\end{figure}
\newpage
\begin{figure}[ht]
\begin{center}
\hspace*{-2.0cm}
\epsfig{
file=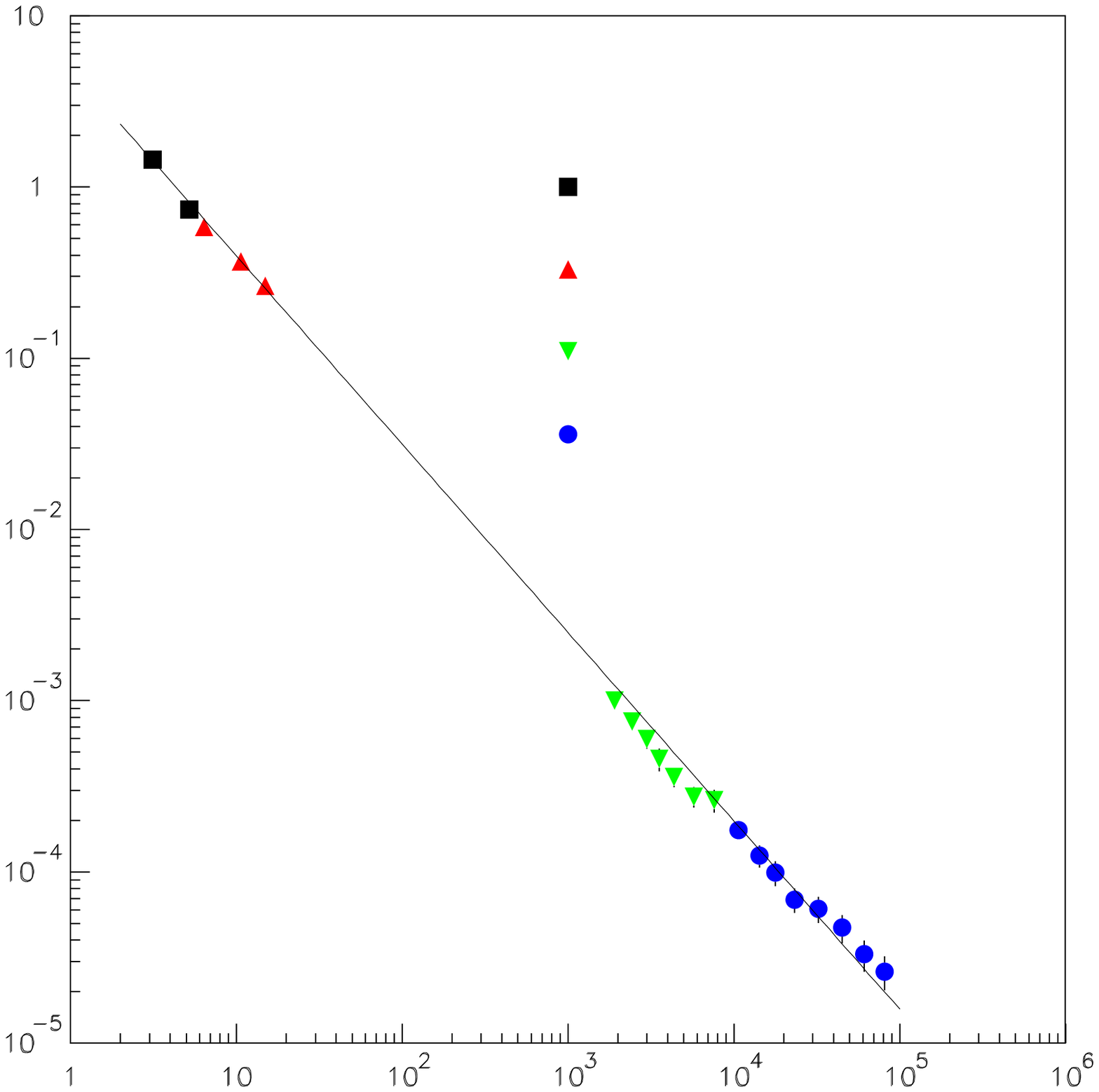,
        height=15.0cm,angle=0}
\put(-50,125){\Large\bf $\sqrt{s}$}
\put(-56,114.5){\Large\bf $14~GeV$}
\put(-56,104){\Large\bf $20~GeV$}
\put(-56,94.5){\Large\bf $546~GeV$}
\put(-56,85){\Large\bf $1800~GeV$}
\put(-100,60){\Large\bf $(\frac{1}{M_X^2})^{1.1}$}
\put(-90,68){\Large\bf\vector(1,1){10}}
\put(-50,0){\Large\bf $M_X^2~[GeV^2]$}
\put(-148,30){\begin{sideways}\huge\bf
$\frac{d^2\sigma}{dtdM_X^2}|_{t=-0.05}[mb/GeV^4]$\end{sideways}}
\put(-71,-35){\bf Figure~5}
\end{center}
\label{Figure-sig-scl}
\end{figure}
\newpage
\begin{figure}[ht]
\begin{center}
\hspace*{-2.0cm}
\epsfig{
file=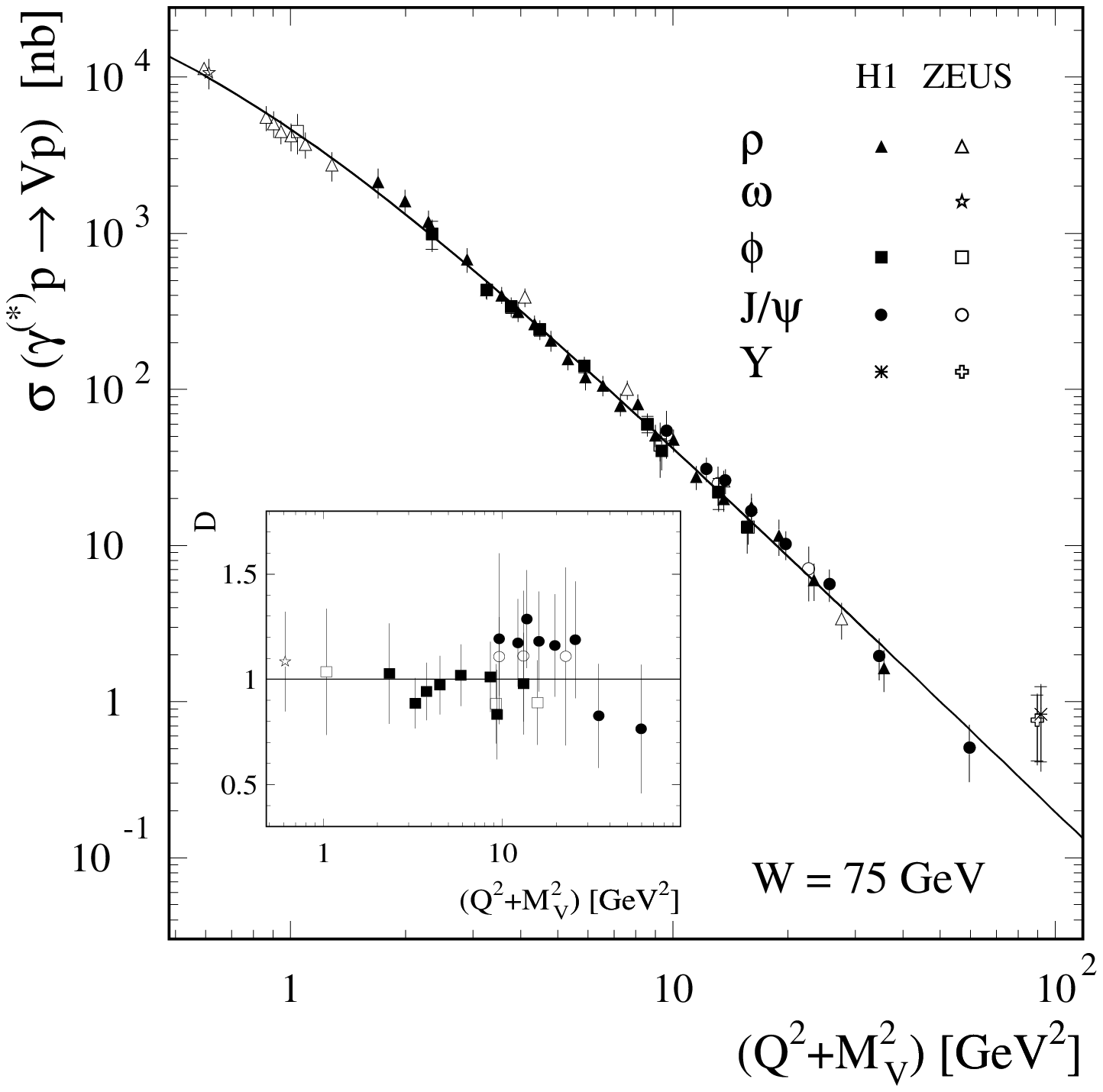,
        bbllx=0,bblly=200,bburx=610,bbury=670,clip=,
        height=15.0cm,angle=0}
\put(-105,-35){\bf Figure~6}
\end{center}
\label{Figure-h1-vm}
\end{figure}
\newpage
\begin{figure}[ht]
\begin{center}
\hspace*{-2.0cm}
\epsfig{
file=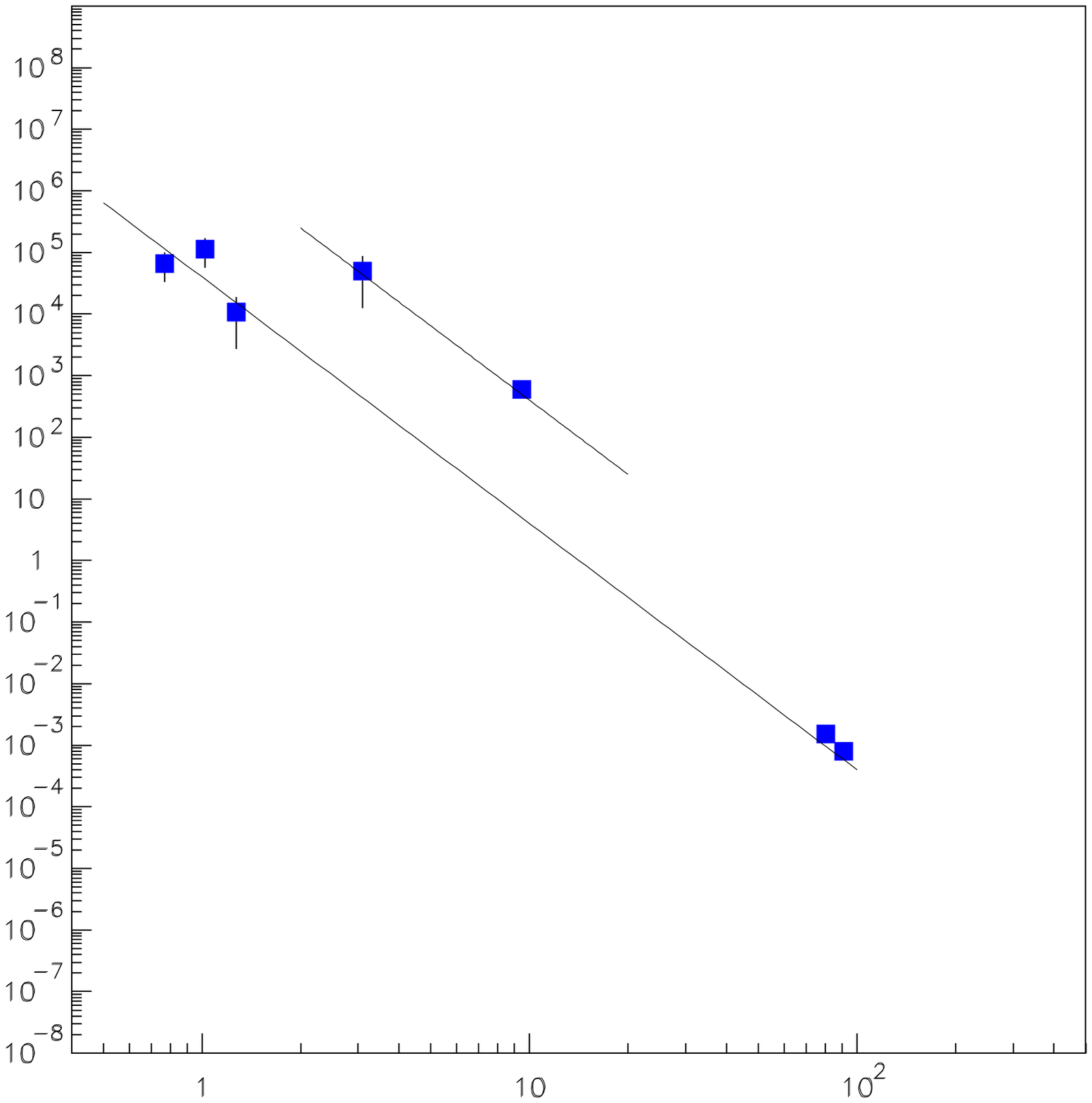,
        height=15.0cm,angle=0}
\put(-112,95){\Large\bf $\rho$}
\put(-105,115){\Large\bf $\phi$}
\put(-103,85){\Large\bf $f_2$}
\put(-90,115){\Large\bf $J/\Psi$}
\put(-68,98){\Large\bf $\Upsilon$}
\put(-38,60){\Large\bf $W^+$}
\put(-35,42){\Large\bf $Z$}
\put(-25,30){\Large\bf $H?$}
\put(-24,43){\color{red}\circle*{25}}
\put(-100,60){\Large\bf $\frac{1}{M^4}$}
\put(-90,68){\Large\bf\vector(1,1){10}}
\put(-50,0){\Large\bf $M~[GeV]$}
\put(-145,60){\begin{sideways}\huge\bf
$\frac{1}{\Gamma}\frac{d\sigma}{dy}(y=0)
[\frac{\mu{b}}{GeV}]$\end{sideways}}
\put(-72,-35){\bf Figure~7}
\end{center}
\label{Figure-mesons}
\end{figure}

\end{document}